\documentclass[11pt,a4paper,twocolumn]{book}
\usepackage{jnpcs}
\usepackage{graphics}
\usepackage{epsfig}
\usepackage{epstopdf}

\ntitle{Human society under the COVID-19 as a complex system. Mean first-passage time methodology}

\nauthor{Anatolii~V.~Mokshin$^{1}$, Yana~A.~Shadrina$^{1}$ and Vladimir~G.~Sherputovskiy$^{2}$}

\naddress{$^1$Kazan Federal University,
      \it 420008 Kazan, Russia \\
   $^2$Republican Medical Information and Analytical Centre of the Republic of Tatarstan,
   \it 420073 Kazan, Russia \\
   {\small \rm E-mail:  anatolii.mokshin@mail.ru}}

\ndata{\today}

\nabstract{Human society under the COVID-19 pandemic can be viewed as a complex system, the evolution of which is characterized by such the parameters as the number of newly diseased, the number of seriously ill patients, the number of those who were identified as diseased, etc. To analyze such a complex system, we propose to apply the methodology known in statistical physics as the methodology of the \emph {mean first-passage times}. As will be shown, this methodology makes it possible to determine properly the so-called threshold, at which a spreading viral disease goes into an epidemic regime with accelerated progression. The efficiency of the methodology is demonstrated by the example of data analysis on the spread of the COVID-19 in  the Russian Federation as well as separately in one of the regions of  the Russian Federation. The methodology can serve as an additional useful tool for solving optimization problems.}

\nkeywords{complex system, mean first-passage times, infectious diseases, critical values, disease patterns}

\begin{document}

\DeclareGraphicsExtensions{.jpg,.pdf,.mps,.png} \firstpage{1}
\nlpage{1} \nvolume{} \nnumber{} \nyear{2023}
\def\nfpage{\thepage}
\thispagestyle{myheadings} \npcstitle

\section{Indroduction \label{sec: introduction}}

The definition of ``\emph{complex system}'' is very subjective in its common usage: each individual (person) can imply and put something of his own into this definition. At the same time, the concept of ``\emph{complexity}'' can be associated with different aspects of a considered system: its internal structure, form, reaction to external disturbances, unpredictability of its behaviour (evolution), impossibility or complexity of its control. Currently,  in natural scientific researches, the definition of ``\emph{complex system}'' is becoming one of the basic definitions in the context of characterizing systems of a certain type~\cite{tempbib1}. The nature of these systems can be very diverse: these can be physical, biological, economic, social, physiological, etc. Despite the fact that the definition of ``\emph{complex system}'' is widely used in the scientific vocabulary and there is a substantive field of science dedicated to the study of complex systems, there is still no universally accepted definition. A discussion of this issue can be found, for example, in the papers~\cite{tempbib2,tempbib3,tempbib4}. Thus, nine completely different definitions are given in Ref.~\cite{tempbib2}. Complex systems are assumed to have the following features. First, any aspect or set of aspects of a complex system is inherently \emph{nonlinear}. Mathematically, it means that the structure, form, dynamics of a considered system, interaction between its constituent elements can be described, in particular, by nonlinear differential equations. Second, a complex system is, generally, \emph{non-Hamiltonian}. In other words, it is difficult or impossible to determine unambiguously  the total energy of such a system. Third, such a system is characterized by \emph{non-integrability} of features of its elements. This means that the sum of the properties of individual components does not necessarily determine the properties of system as a whole. In addition, a typical property of such a system is its \emph{ability to self-organize}, that is expressed in the fact that the system can maintain a certain resistance to external perturbations~\cite{tempbib5,tempbib6}. The aforementioned features are probably the basic features of complex systems of any nature.

In this study, we assume that a complex system is a system that can be \emph{accurately described} only by an infinitely large number of degrees of freedom (factors, parameters)~\cite{AVM_VVM_2019,tempbib7}. On the one hand, such a definition does not contradict the features given above. It also provides a basis for applying the methodology of statistical physics to the analysis and/or description of a complex system. Finally, this definition best reflects the fact that the appropriate tools for studying a complex system are methods of metaheuristics and machine learning (artificial neural networks, genetic algorithms, etc.) that operate with sets of various factors and establish links between these factors.

According to the definition we have given, human society, evolving on its own or under any conditions, can be seen as a complex system. For example, a human society in the face of the spread of an infectious disease (ID) will be determined by an infinitely large number of factors related to demography, migration, climate/weather, economy, level of development of its health care system, and the restrictive regulations in the region concerned. Thus, the character of the spreading ID depends on all these factors to some degree. Some of these factors are the most important, while the impact of other factors may be insignificant. On the other hand, it is possible to focus on studying only one \emph{key} factor that characterizes directly the spread of an ID. In this particular case, it is appropriate to use the number of confirmed cases currently identified as a factor. By analyzing the dynamics of this factor, one can try to determine a trend of ID development, its peculiarities, as well as to make a probabilistic forecast of further ID development, to estimate the probability of ID transition to epidemic and/or pandemic regime~\cite{tempbib8,tempbib9}. Currently, the powerful mathematical methods as well as the machine learning tools are widely used to analyze evolution of ID's~\cite{tempbib27, tempbib10,tempbib11,tempbib12,tempbib13,tempbib14,tempbib15,tempbib26}.

There is the specific feature of such a complex system as human society under conditions of spread of ID. Namely,  when ID is absent or the degree of its development has not reached some threshold at which an epidemic/pandemic regime emerges, it can be assumed that the system is in some stable state. Then, the emergence of an epidemic regime will correspond to escape of the system from this stable state. Here, there is an analogy with the so-called activation-type processes that are well known in physics, chemistry, engineering and biology~\cite{tempbib16,tempbib17,tempbib18}. In this study, it will be shown that the analysis of the spreading ID, which can go into epidemic or pandemic regimes in human society, can be carried out using the so called \emph{mean first-passage time} (MFPT) method well known in modern statistical physics~\cite{tempbib19,tempbib20}. Quantitative results are demonstrated by the example of the COVID-19 spread in the Russian Federation (RF) and in a specific region of the RF.

\section{Methodology \label{sec: method}}
Suppose there is a complex system whose evolution is tracked by means of the time-dependent parameter $n(t)$. The system is assumed to have two stationary (quasi-stationary) states where it can exist and in which the parameter $n$ takes the values $n^{(1)}$ and $n^{(2)}$, respectively. Therefore, we can say that its evolution takes place on a hypothetical free energy landscape $U(n)$, where the values $n^{(1)}$ and $n^{(2)}$ define the positions of the minima. Then, the maximum value of the quantity $U(n)$ enclosed between these minima is determined by the conditions of transition from one state to another, as shown in Fig. 1b.
\begin{figure*}
    \centering
    \includegraphics [clip,width=180mm]{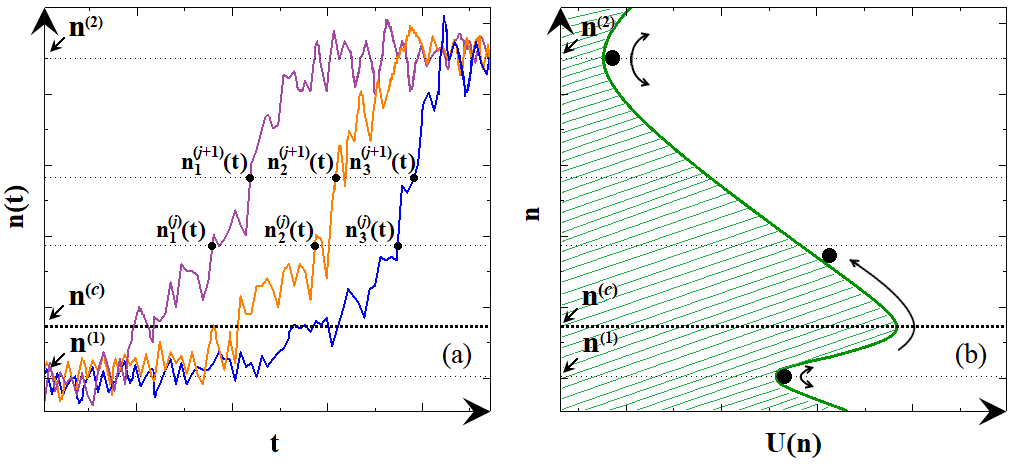}
    \caption{Schematic drawings showing the transition of the system from a local minimum, where the order parameter $n$ takes the value $n^{(1)}$, to another local minimum with the value $n^{(2)}$. (a) Three independent trajectories of the order parameter $n(t)$. The horizontal dotted lines correspond to the values of the order parameters $n^{(1)}$, $n^{(c)}$, $n^{(j)}$, $n^{(j+1)}$, $n^{(2)}$. (b) Free energy landscape $U(n)$, where the quantity $n$ corresponds the abscissa axis, whereas the quantity $U(n)$ is given on the ordinate axis.}
    \label{fig: trajectories_barrier}
\end{figure*}
Here, the quantity $n(t)$ takes the meaning of the reaction coordinate or order parameter characterizing the transition between states. It should be noted that the question of the nature of the free energy $U(n)$ is not fundamental in this case. In physics, this situation corresponds to a two-level system whose evolution happens in a double-well potential~\cite{tempbib16}. The solution corresponding to this situation is used, for example, in the well-known physical problem of a particle overcoming an energy barrier during the transition from one local energy minimum at $n=n^{(1)}$ to another one at $n=n^{(2)}$.

Stay of the system in each of the minima $U(n^{(1)})$ and $U(n^{(2)})$ is determined by the time scales $\tau^{(1)}$ and $\tau^{(2)}$, and the mean (most probable) time to reach the maximum $U(n^{(c)})$ is given by the quantity $\tau^{(c)}$. Thus, $\tau^{(c)}$ determines the most probable time at which the system reaches the critical value $n^{(c)}$, and the inverse quantity $1/\tau^{(c)}$ characterizes the transition rate. The process of transition from one minimum to another is stochastic and is driven by some forces, the nature of which is not crucial in the context of our topic. However, if the transition  process has the character of an activation-type, the energy brought by these forces should be much smaller than $U(n^{(c)})$. For example, from the point of view of classical thermodynamics, the transition between different aggregate states is due to the thermal energy $k_BT$, where $k_B$ is the Boltzmann’s constant and $T$ is the temperature.

The characterization and probabilistic description of such a system can be performed through statistical consideration of various partial realizations, from which we obtain a set of $M$ independent trajectories:
\[
\left\{n_{1}(t),\ n_{2}(t),\ \ldots,n_{i}(t),\ \ldots,\ n_{M}(t)\right\}.
\]
The lower index $i$ is the label of the $i$th realization. This set can be analyzed using the procedure of \emph{inverted averaging}~\cite{tempbib22,tempbib23,tempbib24,tempbib25}, which is based on the MFPT method~\cite{tempbib19}.

The main idea of the MFPT method is to determine the sequence of the first passage times for all possible values of the parameter $n$ for each $i$th trajectory, i.e.:
\[
n_{i}^{(1)}(t),\ n_{i}^{(2)}(t),\ \ldots,n_{i}^{(j)}(t),\ldots
\]
The upper index $(j)$ defines the $j$th value of parameter $n$. In other words, for each $i$th trajectory, $t(n_{i}^{(j)})$ is defined that has the form of a non-decreasing step function (for details, see Fig. 1). Thus, we find the distribution of the average first occurrence time by inverted averaging over the results for $M$ realisations:
\begin{equation} \label{2.1}
  \tau(n)=\frac{1}{M}\sum_{i=1}^{M}t(n_{i}^{(j)}).
\end{equation}

For activation processes, the MFPT curve has a sigmoidal shape.  The larger the ratio of the time scale $\tau^{(c)}$ to the time scale $\tau^{(1)}$, i.e. $\tau^{(c)}/\tau^{(1)} \gg 1$, the steeper the bend will be characterized by this curve. The more pronounced and higher is the energy barrier, the more smooth dependence characterizes the curve $tau(n)$ in the plateau regime (see Fig. 2a).
\begin{figure}
    \centering
    \includegraphics[clip,width=80mm]{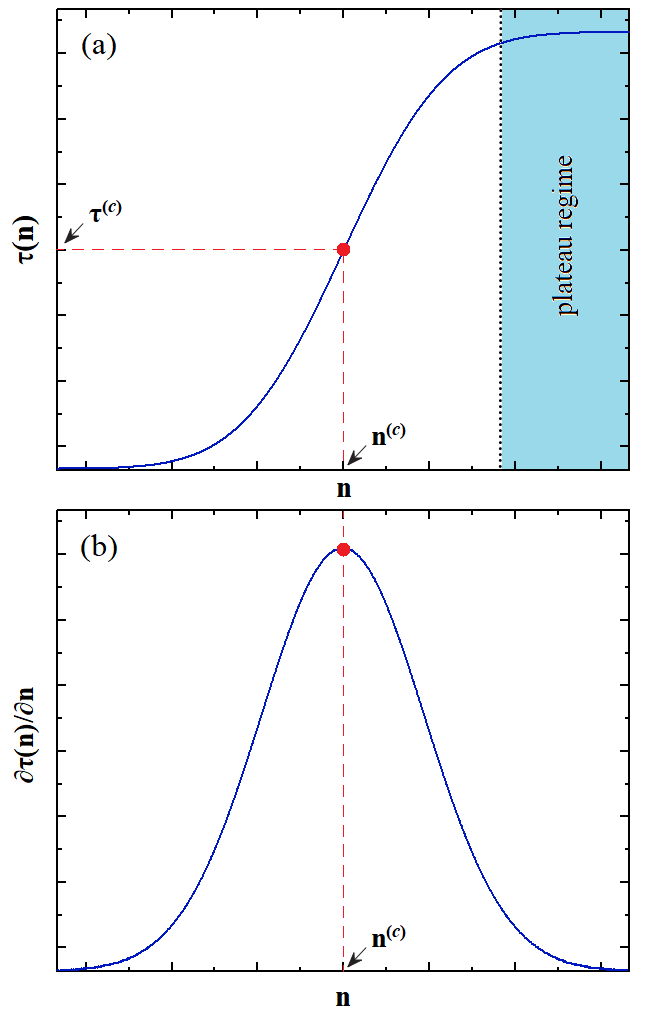}
    \caption{(a) MFPT curve obtained by averaging the different trajectories. The position of the red point on the curve corresponds to the critical value $n^{(c)}$ and the time scale $\tau^{(c)}$. (b) Derivative of the MFPT curve $\partial \tau(n)/\partial n$ whose maximum position corresponds to the critical value $n^{(c)}$.}
    \label{fig: MFPT_base}
\end{figure}
Finally, the position of the first inflection point of the MFPT curve corresponds to the critical value $n^{(c)}$ and the transition time $\tau^{(c)}$ (see Fig. 2). Both of these quantities can be accurately determined by the position of the maximum on the graph of the derivative $\partial \tau(n)/\partial n$ (see Fig. 2b).
Thus, the MFPT methodology allows us to accurately determine the critical value of the order parameter $n$ corresponding to a transition from one state to another and the time scale $\tau$ that characterizes this transition. The applicability of this methodology is not limited to any particular nature of the system. The analysis by means of this technique is based only on a set of independent trajectories for the order parameter $n$, and the number of these trajectories should be enough to obtain a reliable probabilistic estimate.

\section{Description of data \label{sec: Data}}

This paper performs an analysis of daily data describing the spread of the COVID-19 for the period from March $12$, $2020$ to November $16$, $2022$. Two situations are considered: (1) the RF under the COVID-19 and (2) the Republic of Tatarstan (RT), as a separate entity of the RF, also under the COVID-19. Six separate waves of ID were captured during this time period. Statistical analysis has been performed separately for each wave. The parameter $n$ is a quantity that characterizes the number of new confirmed cases of the disease per population of the territorial unit. Thus, quantity $n$ denotes the fraction of diseased persons from the total population for the considered territorial unit.

(1) \emph{The Russian Federation}. -- The analysis is based on the data for $72$ regions of the RF (see Fig. 3a).
\begin{figure*}
    \centering
    \includegraphics [clip,width=160mm]{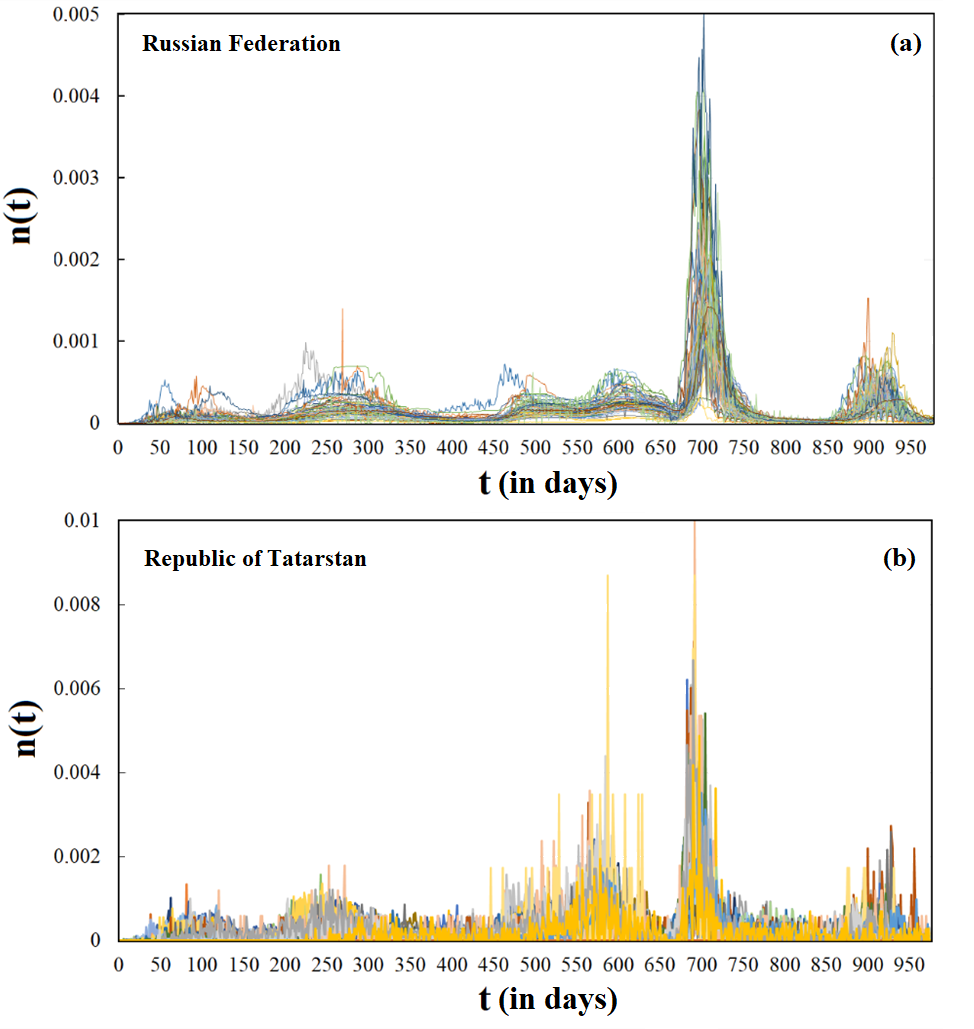}
    \caption{Dynamics of the COVID-19 incidence in the RF and the RT for the period from March $12$, $2020$ to November $16$, $2022$. Individual trajectories correspond to different regions in the case of the RF and different health facilities in the case of the RT (for details, see the text).}
    \label{fig: Covid_RF_RT}
\end{figure*}
The data for the RF is taken from the official portal ``stopcoronavirus.rf''~\footnote{https://xn--80aesfpebagmfblc0a.xn--p1ai/}, where information on disease incidence is updated daily. Various regions were considered, including the largest, such as Moscow, the Moscow Region and St. Petersburg, the average populations of which are $12\,678\,079$, $7\,690\,863$ and $5\,398\,064$ people, respectively. The average population of such small regions as the Nenets Autonomous Area, the Chukotka Autonomous Area and the Jewish Autonomous Region for the period under review was $44\,111$, $50\,288$ and $158\,305$ people, respectively. The population density of the regions of the RF considered in the given study ranges from $0.1$ to $4950.4$ people/km$^2$. The regions with the lowest population density are the Chukotka Autonomous Area ($0.1$ people/km$^2$), the Nenets Autonomous Area ($0.2$ people/km$^2$) and the Republic of Sakha (Yakutia) ($0.3$ people/km$^2$). The highest densities are in such the regions as Moscow ($4\,950.4$ people/km$^2$), St. Petersburg ($3\,858.5$ people/km$^2$) and the Moscow Region ($173.5$ people/km$^2$). The data source is the Federal State Statistics Service for 2020~\footnote{https://rosstat.gov.ru/folder/210/document/13205}.

(2) \emph{The Republic of Tatarstan}. -- The study considers data for $97$ health facilities of the RT (see Fig. 3b), where confirmed cases of the disease were registered. The values of the number of diseased persons were recalculated taking into account the population attached to the respective health facility \footnote{Population of municipalities of the Republic of Tatarstan at the beginning of 2020. Statistical Bulletin.}$^{,}$\footnote{https://msu.tatarstan.ru/mregions.htm}. The data were relevant to both large settlements with several health facilities represented and those with small numbers of people living in them.

The period taken into consideration for both the systems was $980$~days. As shown in Fig.~\ref{fig: Covid_RF_RT}, the dynamics of daily detected cases both nationally (Fig.~\ref{fig: Covid_RF_RT}a) and regionally (Fig.~\ref{fig: Covid_RF_RT}b) for a given time period reveals six spikes/waves where there is a sharp increase in prevalence. It should be noted that it is practically impossible to determine the boundaries between individual waves with sufficient accuracy from the original data. In this case, the third and fourth waves practically merge.

\section{Results for the Russian Federation \label{sec: Results}}

\subsection{The start of the next wave of the epidemic} \label{Sec: Start_RF}

The key point of the analysis is to determine the time moment that can be taken as the beginning of the next wave. The first wave was excluded from our analysis because the information on morbidity at the initial stages of the COVID-19 was inconsistent and incomplete.

\begin{figure}
\centering
    \includegraphics [clip,width=90mm]{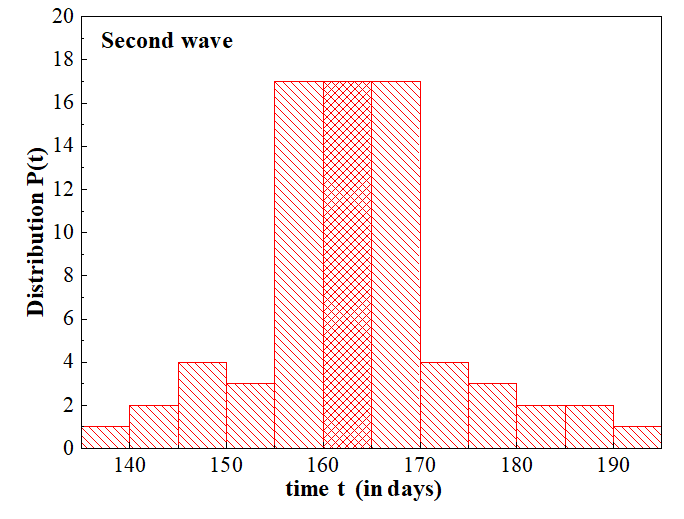}
    \caption{Distribution of times of minimum number of cases in different regions of the RF before the second wave of the COVID-19.}
    \label{fig: Distr_times}
\end{figure}

An accurate statistical estimate of the time at which the next wave begins can be made on the basis of the distribution of times when the number of cases in the regions was minimal. The position of the maximum on this distribution can be taken as the time of the beginning of the next wave of the COVID-19. Thus, Fig.~\ref{fig: Distr_times} shows such a distribution for the stage corresponding to the beginning of the second wave of ID progression.
The time step of the distribution (i.e. bin width) is $5$~days. This quantity determines the size of the confidence interval around the resulting estimates for the beginning date of the corresponding wave. In addition, this time step appears to be optimal for reasons of the statistics used to construct the distribution based on data for~$100$ independent objects, as well as for reasons of the resulting ratio between disease wave size and time step. From this distribution, we find that the date corresponding to $162.5$ ($\pm 2.5$) days from the beginning of the COVID-19 leakage registration can be chosen as the initial moment of the second wave. Similar probabilistic estimates were performed for all other waves of the COVID-19. Thus, for the third wave we find that its onset is 427.5 days, for the fourth wave -- $547.5$ days, for the fifth and sixth waves -- $667.5$ and $837.5$ days, respectively.

Based on the results obtained, the full time interval of $980$ days was divided into fragments corresponding to individual disease waves. Subsequent analysis was performed separately for each wave of the COVID-19 (except for the first wave).

\subsection{Critical values of the transition to the epidemic regime for different waves}

\begin{figure*}
    \centering
    \includegraphics [clip,width=180mm]{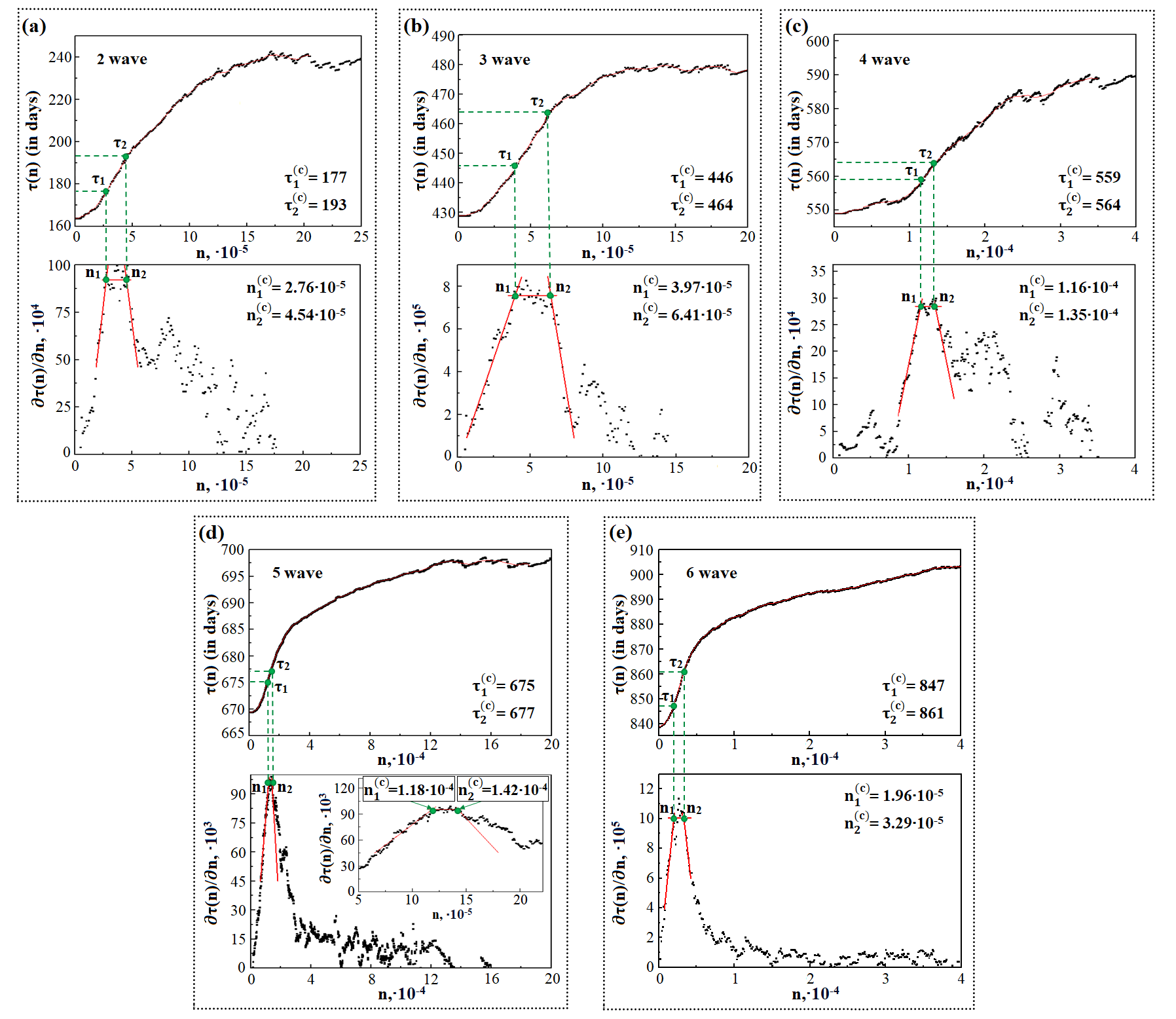}
    \caption{MFPT $\tau(n)$ curves and derivatives $\partial \tau(n)/\partial n$ determined for different disease waves. Here, $\tau_{1}^{(c)}$ and $\tau_{2}^{(c)}$ define the critical time range, $n_{1}^{(c)}$ and $n_{2}^{(c)}$ define the boundaries of the parameter $n$ with the critical values corresponding to this range. Insert in panel (d): enlarged fragment of the main graph.}
    \label{fig: MFPT_derivatives}
\end{figure*}

The time distributions of $\tau(n)$  for each wave were obtained using the MFPT analysis. The obtained results are shown in Fig.~\ref{fig: MFPT_derivatives}.
In the case when the variable $n$ represents a characteristic of the spread of ID - the daily number of newly infected people, the quantity of $\tau$ will make sense of the time moment when the number of infected people $n$ takes some specific value.  As can be seen from the shapes of the obtained distributions, the evolution of the next wave follows the scenario typical for activation-type processes. This is indicated by the fact that all the distributions have a pronounced sigmoidal shape, similar to one in Fig.~\ref{fig: MFPT_base}a. Thus, the smooth form of each obtained distribution of $\tau(n)$ for the initial stage of the disease wave, is replaced by a sharp increase, which, in turn, passes into a saturation regime (plateau).

It turns out to be very interesting that the inflection point in each distribution $\tau(n)$, defined by the critical values of $\tau(n)$ and $n^{(c)}$, is not characterized by single values of these quantities, but by ranges of values $\tau^{(c)}\in[\tau_{1}^{(c)};\tau_{2}^{(c)}]$ and $n^{(c)}\in[n_{1}^{(c)};n_{2}^{(c)}]$. The determined critical values are given in Tab. \ref{tab: parameters_RF}.
\begin{table*}[ht]
    \caption{Parameters characterizing the course of the different the COVID-19 waves in the RF: start date of the $i$th wave, ranges of values of the critical time $\tau^{(c)}$ (in days) and the parameter $n^{(c)}$ for the $i$th wave, average peak characteristics of the $i$th wave: $\bar{\tau}_{i}^{(max)}$ and $\bar{n}_{i}^{(max)}$.}
	\label{tab: parameters_RF}
        \centering
	\begin{tabular}{@{}cccccc@{}}
		\hline
		\hline
        $i$th wave & start date & $\tau_{i}^{(c)}$  & $n_{i}^{(c)},\cdot10^{-5}$ & $\bar{\tau}_{i}^{(max)}$ & $\bar{n}_{i}^{(max)}, \cdot10^{-4}$ \\
		\hline
        \hline
		2 & 162.5$\pm$2.5 (20.08.2020) &  [14; 30] & [2.76; 4.54] & 117 & 2.29 \\
		3 & 427.5$\pm$2.5 (13.05.2021) &  [18; 36] & [3.97; 6.41] & 81 & 1.98 \\
		4 & 547.5$\pm$2.5 (10.09.2021) &  [11; 16] & [11.6; 13.5] & 63 & 2.95 \\
		5 & 667.5$\pm$2.5 (08.01.2022) &  [7; 9] & [11.8; 14.2] & 37 & 16.7 \\
		6 & 837.5$\pm$2.5 (27.06.2022) &  [9; 23] & [1.96; 3.29] & 82 & 4.71 \\
 		\hline
 		\hline
	\end{tabular}
\end{table*}
This result seems rather unexpected, although understandable. It indicates that the onset of an irreversible increase in the number of new cases falls within a time range of a few days to three weeks, depending on the particular wave. Thus, this range is the largest in the case of the third wave of the disease.  It also turns out that the critical proportion of all confirmed cases of the disease, the excess of which correlates with the explosive spread of the disease, is about $1\cdot10^{-5}$ of the total population in the corresponding region. For example, in the case of the second wave for Moscow, we find that the absolute critical values of the confirmed cases of the disease are in the range [$350$; $576$] people, and for the Nenets Autonomous Area we get the range [$2$; $3$] people. These absolute critical values are quite commensurable if we take into account the total population living in the respective regions. The obtained values clearly reflect the fact that the COVID-19 belongs to extremely highly ID's.

\subsection{Results for the Republic of Tatarstan as a region of the Russian Federation}

In the case of the data for the RT, the time moments of the start of each disease wave were estimated using the same scheme as for the RF (see description \ref{Sec: Start_RF}). The values of the time moments were obtained which were found to correlate with the results for the RF (see Tab. \ref{tab: RT}) were obtained.
\begin{table*}[ht]
    \caption{Parameters characterizing the course of different waves of the COVID-19 in the RT (by analogy with Tab. 1).}
	\label{tab: RT}
        \centering
	\begin{tabular}{@{}cccccc@{}}
		\hline
		\hline
        $i$th wave & start date & $\tau_{i}^{(c)}$  & $n_{i}^{(c)},\cdot10^{-5}$ & $\bar{\tau}_{i}^{(max)}$ & $\bar{n}_{i}^{(max)}, \cdot10^{-4}$ \\
		\hline
        \hline
		2 & 176$\pm$2 (03.09.2020) &  [18; 26] & [0.49; 0.85] & 117 & 5.24 \\
		3 & 426$\pm$2 (11.05.2021) &  [27; 35] & [1.59; 2.07] & 104 & 6.73 \\
		4 & 550$\pm$2 (12.09.2021) &  [8; 13] & [4.77; 6.99] & 36 & 13.4 \\
		5 & 656$\pm$2 (27.12.2021) &  [6; 11] & [0.58; 1.27] & 36 & 34 \\
		6 & 856$\pm$2 (15.07.2022) &  [7; 17] & [0.55; 1.47] & 58 & 7.23 \\
 		\hline
 		\hline
	\end{tabular}
\end{table*}

The obtained MFPT-curves also indicate the activation character of the progression of each wave of the COVID-19 in the RT. Analysis of these curves reveals that there are the ranges of critical values of $\tau^{(c)}$ and $n^{(c)}$ that generally correlate with the results obtained for the country as a whole.

\subsection{Epidemic patterns} \label{Sec: Epid_patterns}

\begin{figure}
    \centering
    \includegraphics [clip,width=90mm]{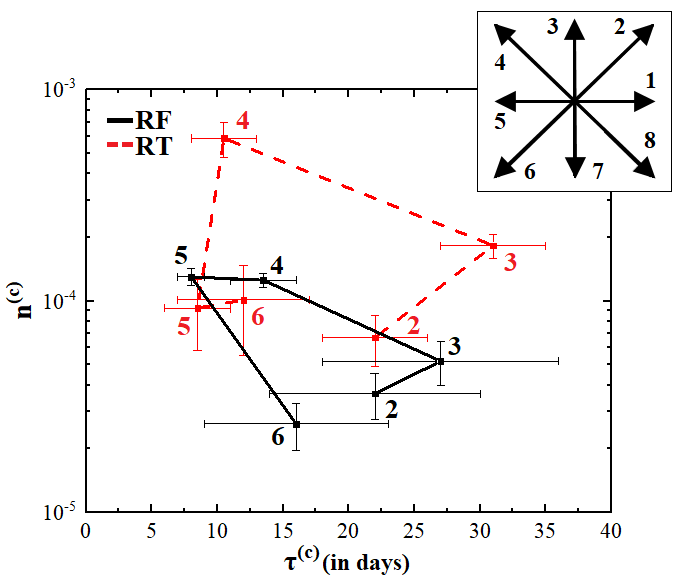}
    \caption{
    Patterns of the critical values of $\tau^{(c)}$ and $n^{(c)}$ of individual waves determined for the second ($i=2$), third ($i=3$), $\ldots$ and sixth ($i=6$) waves of the COVID-19 in the RF (solid black line) and in the RT (dashed red line). The value of $i$ indicates the order number of the considered wave. \textbf{Insert}: The main directions of the trajectory in the pattern according to the development of the epidemic (see Sec.~\ref{Sec: Epid_patterns}).}
    \label{fig: Pattern_crit_values}
\end{figure}

A common feature of most ID's is the occurrence of separate waves. At the same time, the course of diseases is largely determined by the way these waves evolve, i.e. how the disease spreads at the stage of separate waves. In this regard, it is useful to consider the so-called \textit{disease patterns}, which represent the correspondence of different characteristics for a sequence of waves. Thus, Fig.~\ref{fig: Pattern_crit_values} shows the patterns for the critical wave characteristics -- the critical value $n_{i}^{(c)}$ and the critical time $\tau_{i}^{(c)}$.
In the figure, the abscissa axis represents the time characteristics and the ordinate axis displays the parameters characterizing the fractions of infected individuals. As the ID progresses and moves from one wave to the next, some directional trajectories will emerge in these patterns.

For such the patterns, the following main scenarios can be highlighted. The upward direction of the trajectory (labelled \textbf{``direction $3$''}) implies an increase in the critical value of the subsequent wave. This scenario is possible if the system adapts to an ID and a virus mutates accordingly. In this case, the transition of the system into the epidemic regime at the stage of the next wave is possible only at a larger critical value of $n^{(c)}$ compared to the critical value of the previous wave. Here, the adaptation of the system may be largely due to the actions implemented, as the system is ``trained'' to cope with the challenges caused by the ID. On the other hand, the downward direction of the trajectory (\textbf{``direction $5$''}) may indicate mutations in the virus that make it more infectious, as well as a weakening of the system’s ``immunity''. The direction of the trajectory to the right side of the pattern (\textbf{``direction $1$''}) indicates that the value of the critical time $\tau$ of the next wave exceeds the
critical value of the previous wave. This is evidence that the spread of the ID is slowing down, which may be due to both mutations in the virus as well as the system’s response to it. On the other hand, the direction of the trajectory to the left (\textbf{``direction $7$''}) reflects the fact that the ID is becoming more contagious and the virus is spreading faster in the system. Explanations are provided in insert to Fig.~\ref{fig: Pattern_crit_values}.

The diagonal directions of the pattern will be interpreted as follows:
\begin{enumerate}
\item[$\bullet$] The right and up (\textbf{``direction $2$''}). This trajectory direction corresponds to a situation where a virus is present and active, but from wave to wave the system is trained to deal with it more successfully.
\item[$\bullet$] The left and up (\textbf{``direction $4$''}).  This direction of the trajectory represents the case when the infectivity of the virus becomes higher as it progresses to the next wave, but the system, as a whole, copes with the virus.
\item[$\bullet$] The left and down (\textbf{``direction $6$''}). This is possible when the virus is ``successful''. This is the worst-case scenario for the spread of an ID in the system and can lead to a fatal ending.
\item[$\bullet$] The right and down (\textbf{``direction $8$''}). In this case, the virus remains highly contagious. However, specific conditions are required for this contagion to be realized. This situation is possible when the system successfully fights a virus and defeats it.
\end{enumerate}

On the basis of this, it becomes clear that for a young, active virus, the initial stages of its progression and spread will be associated with movement along the directions $2 \rightarrow 3 \rightarrow 4$. On the other hand, the implementation of ``direction $8$'' may indicate the finality of virus activity and may be due to virus degeneration.

It can be seen in Fig.~\ref{fig: Pattern_crit_values} that the initial stages of the trajectories related to the $2$nd, $3$rd and $4$th waves for the RF and the RT proceed in a similar way.  Moreover, the character of the direction of these trajectories is typical for a young ID such as the COVID-19. Further, the transition from the $4$th wave to the $5$th wave occurs in the directions (``direction $7$'' for the RF and ``direction $5$'' for the RT), which are typical for the cases when an ID becomes highly contagious. This situation was largely due to  the spread of the ``Omicron'' strain. However, the high infectiousness of the ``Omicron'' strain was compensated for by its milder course. As a result, the further spread of the ID, corresponding to the transition from the $5$th wave to the $6$th wave, began to evolve according to ``direction $8$'' for the RF and according to ``direction $1$'' for the RT,  which are associated with a significant decrease in virus activity and a slowing of the spread of the ID.

\begin{figure}
    \centering
    \includegraphics [clip,width=90mm]{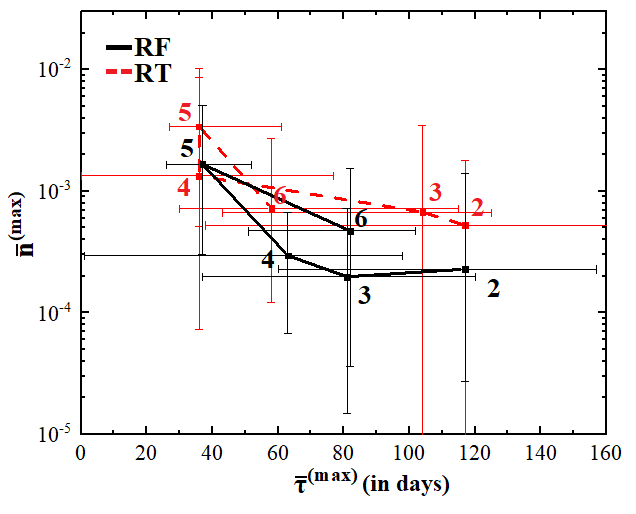}%
    \caption{Patterns of peak values of $\bar{\tau}^{(max)}$ and $\bar{n}^{(max)}$ of individual waves determined for the second, third, $\ldots$ and sixth waves of the COVID-19 in the RF (solid black line) and in the RT (dashed red line).}
    \label{fig: patterns_peaks}
\end{figure}

The obtained results do indicate that with the emergence of the ``Omicron'' strain, the character of the spread of the COVID-19 is changing dramatically. At the same time, the results for the RT indicate that there is a certain preparatory stage for these changes, which takes place during the transition from the $4$th to the $5$th wave. The same stage is clearly manifested in the direction of the trajectory, which characterizes the attainment of peak values of $\bar{\tau}^{(max)}$ and $\bar{n}^{(max)}$ by separate waves (see Fig. \ref{fig: patterns_peaks}).

Here $\bar{n}^{(max)}$ is the fraction of the total population that is ill when the wave attains its peak; and $\bar{\tau}^{(max)}$ is the time at which an individual wave peaks (averaged values of both the parameters are considered). In the case of the RF, we have a trajectory typical for a situation, when the activity of the virus increases from second wave to fifth wave; and only the last stage displays a decrease in its activity. For the RT, a separate stage is distinguished at the transition from the $4$th to the $5$th wave, where the trajectory is significantly different from what it was before. The trajectories at the last stage have a completely similar character.

\section{Conclusion}

We conclude with the following main results.
\begin{enumerate}
\item[(i)] It is shown that the MFPT method can be used to analyze the evolution of a specific complex system such as human society under conditions of an ID spread.
\item[(ii)] The COVID-19 propagation in the RF as well as in the RT is an activation-type process.
\item[(iii)] It is found that the critical values of $\tau^{(c)}$ and $n^{(c)}$ associated with the transition of the ID to the epidemic regime are given by certain ranges of values that are accurately determined by the MFPT method.
\item[(iv)] The results obtained for a single region (RT) were found to be fully correlated with the results for the country (RF) as a whole.
\item[(v)] We propose a methodology for analyzing the development of ID based on a detailed examination of the patterns that represent the correspondence between the critical characteristics for a series of ID waves.
\item[(vi)] Based on the character of the obtained patterns, it was directly found that the COVID-19 spreads in both the RF and the RT according to a scenario typical for a young ID.
\end{enumerate}


\end{document}